\documentclass[12pt,preprint]{aastex}

\slugcomment{Accepted for publication in ApJ}

\shorttitle{DH Type II Radio Burst and CME-CME Interaction}
\shortauthors{M\"akel\"a et al.}

\begin{document}

\title{Source Regions of the Type II Radio Burst Observed During a CME-CME Interaction on 2013 May 22}

\author{P. M\"akel\"a\altaffilmark{1}}
\affil{The Catholic University of America, Washington, DC 20064, USA}
\email{pertti.makela@nasa.gov}

\author{N. Gopalswamy}
\affil{NASA Goddard Space Flight Center, Greenbelt, MD 20771, USA}

\author{M. J. Reiner\altaffilmark{1}, S. Akiyama\altaffilmark{1}}
\affil{The Catholic University of America, Washington, DC 20064, USA}

\and

\author{V. Krupar\altaffilmark{2}}
\affil{The Blackett Laboratory, Imperial College London, SW7 2AZ London, UK}

\altaffiltext{1}{NASA Goddard Space Flight Center, Greenbelt, MD 20771, USA}
\altaffiltext{2}{Institute of Atmospheric Physics CAS, 141 31 Prague, Czech Republic}

\begin{abstract}
We report on our study of radio source regions during the type II radio burst on 2013 May 22 based on direction finding (DF) analysis of the Wind/WAVES and STEREO/WAVES (SWAVES) radio observations at decameter-hectometric (DH) wavelengths. The type II emission showed an enhancement that coincided with interaction of two coronal mass ejections (CMEs) launched in sequence along closely spaced trajectories. The triangulation of the SWAVES source directions posited the ecliptic projections of the radio sources near the line connecting the Sun and the STEREO-A spacecraft. The WAVES and SWAVES source directions revealed shifts in the latitude of the radio source indicating that the spatial location of the dominant source of the type II emission varies during the CME-CME interaction. The WAVES source directions close to 1 MHz frequencies matched the location of the leading edge of the primary CME seen in the images of the LASCO/C3 coronagraph. This correspondence of spatial locations at both wavelengths confirms that the CME-CME interaction region is the source of the type II enhancement. Comparison of radio and white-light observations also showed that at lower frequencies scattering significantly affects radio wave propagation.
\end{abstract}


\section{Introduction}

Coronal mass ejections (CMEs) are magnetized plasma clouds launched from the solar corona into the interplanetary (IP) space. They are significant and possibly hazardous elements of solar activity, because Earth-directed CMEs have potential to cause severe space weather disturbances \citep{webbhoward12}. CMEs propagating through the solar corona and the IP space faster than the local Alfven speed are able to drive a shock ahead of them. These CME-driven shocks can accelerate particles, mostly protons and electrons, into very high speeds. Electrons accelerated by the shock can excite Langmuir waves, which in turn interact with low-frequency waves and generate radio waves at the local plasma frequency or its second harmonic. Resulting radio waves are detected remotely by radio receivers as type II radio bursts, which can be identified as emission features drifting slowly towards the lower frequencies. The rate of the frequency decrease is related to the propagation of the CME-driven shock, because plasma frequency is proportional to the square root of the plasma density, which decreases with the radial distance from the Sun. Hence, the frequency drift decrease of type II emission depends on the gradient of the plasma density and the speed of the CME-driven shock that propagates away from the Sun through the IP medium.

The CME occurrence rate is known to follow the solar cycle activity, changing from one CME every other day during the minimum to 5--6 CMEs per day during the solar maximum \citep{webbhoward94,gopalswamy06}. Occasionally two CMEs are launched from a same active region in a close sequence. If the second CME is faster than the preceding CME, it will catch up to the preceding slower CME resulting in a CME-CME interaction event. \citet{gopalswamyetal01, gopalswamyetal02a} reported the first detection of an enhancement of type II radio emission at decameter-hectometric (DH) wavelengths caused by an interaction between two CMEs in the IP space. \citet{gopalswamyetal02b} found evidence that CME-CME interactions are connected to solar energetic particle (SEP) events, indicating that CME-CME interactions might enhance the acceleration efficiency of both electrons and ions \citep[see also, ][]{gopalswamyetal04,lietal12,dingetal13,shanmugarajuprasanna14}. Recently, there have been multiple observational studies and numerical simulations on CME-CME interactions that concentrate on the kinematics of the CMEs, including reports of three interacting CMEs \citep[see][and references therein]{colaninnovourlidas15}.

The methods of obtaining the arrival direction of an oncoming radio wave are called direction-finding (DF) methods or alternatively goniopolarimetry \citep{cecconi07}. The DF analysis technique depends on the type of the spacecraft attitude stabilization. For receivers on spinning spacecraft, the signal modulation due to the spacecraft spin will be analyzed \citep[see, e.g.,][]{fainbergstone74,manningfainberg80,reinerstone88,reineretal98}. For receivers on stabilized (non-spinning) spacecraft, autocorrelations and cross correlations of the antenna signals will be used in the analysis \citep[see, e.g.,][]{lecacheux78,santoliketal03,cecconizarka05,kruparetal12,martinezoliverosetal12a}. Both techniques exploit the anisotropic beaming pattern of the receiving dipole antennas to estimate the radio source direction. Here we use data both from the WAVES \citep{bougeretetla95} radio experiment on the spinning Wind spacecraft and the SWAVES \citep{bougeretetal08} radio experiment on the three-axis stabilized Solar TErrestrial RElations Observatory (STEREO) spacecraft \citep{kaiseretal08}.

The DF analysis of measurements by one spacecraft can give the direction of the radio source as seen from the observing spacecraft, but to find the spatial location of the radio source one needs to use triangulation of DF measurements from at least two spacecraft that have sufficiently different viewpoints. \citet{liuetal10} have developed a geometric triangulation technique for estimating the CME location from multi-viewpoint white-light images, which can also be applied to the radio DF measurements. One can also estimate the source location by calculating the closest points on the two lines corresponding the source directions as seen from the two observing spacecraft (see, e.g., \url{http://www.geomalgorithms.com/a07-\_distance.html}). The lines corresponding to the radio source directions seen from the two spacecraft normally do not intersect each other. In general, it is known that refraction due to solar wind density gradients and scattering by inhomogeneities in the IP density, while the waves travel from the source to the spacecraft, affect the visibility and apparent location of radio bursts \citep[e.g.,][]{thejappaetal07}. If the radio source extends over a large region or there are multiple radio sources, the observed radio emission will be the sum over the extended source or over the separate sources. In addition for radio receivers at different locations, spatial directivity of the radio burst can affect wherefrom the dominant fraction of the measured emission originates.

We report on DH type II radio observations of a CME-CME interaction event on 2013 May 22. In addition, \citet{dingetal14} have shown that the release times of SEPs detected during this event agreed with the start time of the type II burst. They also fitted the Graduated Cylindrical Shell (GCS) model \citep{thernisien11} to the images of both CMEs, which showed that both CMEs were propagating along the same direction. We estimated the source location of the associated type II enhancement by applying radio DF and triangulation techniques to data from Wind and STEREO radio instruments. Similar radio DF and triangulation methods was previously used to study the CME-CME interaction event on 2010 August 1 by \citet{martinezoliverosetal12b}. They found evidence for a causal relationship between the type II radio burst and the CME-CME interaction. Another recent study by \citet{magdalenicetal14} used DF and triangulation methods to investigate the radio source location of type II radio burst during the 2012 March 5 solar eruption. They found that IP type II emission originated from a source at the southern flank of the CME where the CME-driven shock appeared to be interacting with a coronal streamer. Interaction of a CME with a dense streamer has been considered to have the same effect as CME-CME interaction \citep{gopalswamyetal04}.

\section{The 2013 May 22 Solar Eruptions}

The primary solar eruption on 2013 May 22 was associated with a M5.0 soft  X-ray flare at 13:08 UT from AR11745 located at N15$^{\circ}$W70$^{\circ}$. At 13:25 UT the C2 coronagraph of the Large Angle and Spectrometric Coronagraph \citep[LASCO;][]{brueckneretal95} on the Solar and Heliospheric Observatory (SOHO) spacecraft observed the associated CME with a sky-plane speed of 1466 km~s$^{-1}$ (see \url{http://cdaw.gsfc.nasa.gov/CME\_list/index.html}).  \citet{gopalswamyetal14} estimated that the peak speed of the CME was $\sim$ 1880 km~s$^{-1}$.  The early kinematics of the primary CME has been studied by \citet{chengetal14}. They found that the magnetic flux rope, i.e., the primary CME started accelerate around 12:31 UT. This primary CME was preceded by a CME with a sky-plane speed of 687 km~s$^{-1}$, first detected by LASCO/C2 at 08:48 UT. The preceding CME was associated with a C2.3 flare at N17$^{\circ}$W62$^{\circ}$ that occurred at 08:02 UT in the same AR11745 as the flare associated with the primary CME. Both CMEs were also observed by the coronagraphs and heliospheric imagers of the Sun Earth Connection Coronal and Heliospheric Investigation \citep[SECCHI;][]{howardetal08} on the STEREO spacecraft. At the time of the eruption the heliographic longitudes of STEREO-A and B were W137$^{\circ}$ and E141$^{\circ}$, respectively. Therefore, the SOHO and STEREO-A spacecraft were viewing the CMEs from opposite sides, but with similar side views. From the STEREO-B viewpoint the CME source region was behind the Sun, but the primary CME was clearly visible already at 13:05 UT in the COR1 field of view. Because the LASCO/C3 field of view (FOV) extends out to $\sim$ 32 Rs compared to the FOV of the SECCHI/COR2 which observes up to $\sim$ 15 Rs, the LASCO/C3 coronagraph provided the best view to study the evolution of the CME-CME interaction that occurred in the vicinity of the height of 15 Rs. Around 13:35 UT, the STEREO and Wind radio instruments observed the start of a sporadic DH type II radio burst that intensified approximately one hour later at frequencies above 1 MHz. This enhancement progressed down to $\sim$ 300 kHz frequency level, where it faded temporarily around 16:20 UT (Figure~\ref{fig1}). In addition to the DH type II radio burst, the NOAA Solar Edited Events list reports a metric type II around 12:29 UT observed by the Sagamore Hill radio observatory. Near Earth around 14:20 UT, an SEP event with $>$ 10 MeV proton flux peak of 1660 pfu (1 pfu = 1 proton cm$^{-2}$~sr$^{-1}$~s$^{-1}$) was observed by the GOES satellite.

Figure~\ref{fig1} shows the Wind/WAVES dynamic spectrum together with two running difference images from LASCO/C3. The main enhancement of type II emission is observed to start close to the time of the first C3 image at 13:54 UT (Fig.~\ref{fig1}(a)). The eruption was also associated with a strong type III radio burst starting around 13:10 UT, but here we analyze only the DH type II burst. In the C3 image, the primary CME is seen as a bright feature emerging from behind the occulting disk. The preceding CME is the fainter structure ahead of the primary CME. In the second C3 image (Fig.~\ref{fig1}(b)) at 16:18 UT, the primary and the preceding CMEs are already fully overlapping. At the same time in the WAVES dynamic spectra (Fig.~\ref{fig1}(c)), the main enhancement of type II emission is seen to weaken and also the drift rate of the type II burst appears to change simultaneously. Therefore, we believe that the type II enhancement is a similar radio signature of the CME-CME interaction that was first reported by \citet{gopalswamyetal01}. The dynamic spectra of STEREO A and B in the Figs~\ref{fig1}(d) and (e) reveal similar emission patterns, even though the type II burst observed at STEREO-B had a lower peak intensity and a narrower bandwidth. Only the most intense part of the emission was visible above the background. The weakness of the radio event at STEREO-B may be due in part to the backside location of the solar source as seen from STEREO-B (see the flare and spacecraft locations plotted in Fig.~\ref{fig2}). The backside solar source also explains the lower starting frequency (6 MHz) of the associated type III radio burst at the STEREO-B (Fig.~\ref{fig1}e).

\section{Direction Finding}

In our analysis we have used radio data from the Wind and STEREO spacecraft. The High Frequency Receiver 1 (HFR1) and 2 (HFR2) of STEREO/WAVES (SWAVES) have a combined frequency range extending from 125 kHz to 16 MHz. However, only HFR1 that operates at the frequency range of 125--1975 kHz provides the autocorrelations and cross correlations required for the DF analysis. The center frequencies of the channels increase at 50 kHz steps. The RAD1 and RAD2 receivers of Wind/WAVES have a comparable frequency range of 20 kHz--14 MHz. Measurements at eight frequencies of the RAD1 receiver covering the range 20 kHz -- 1.04 MHz was used for the Wind/WAVES DF analysis. The DF frequencies we used are 428 kHz, 484 kHz, 548 kHz, 624 kHz, 708 kHz, 804 kHz, 916 kHz, and 1040 kHz. In our DF data set only two frequencies (428 kHz and 624 kHz) of the WAVES/RAD1 receiver match closely enough those (425 kHz and 625 kHz) of the STEREO/HFR1 receiver to be useful for triangulation of the spatial location of the radio source. Even though triangulation of DF measurements by all three spacecraft has been used previously \citep[see, e.g.,][]{reineretal09}, we have analyzed the Wind and STEREO measurements separately.

\subsection{STEREO/SWAVES Observations}
Figure~\ref{fig2} shows the results from the DF analysis and triangulation of the STEREO A and B data based on methods by \citet{martinezoliverosetal12a} and \citet{liuetal10}. We estimated the direction of the radio source at the STEREO spacecraft at the time of the maximum intensity of each frequency. This was necessary because the type II signal was much weaker at STEREO-B. In general, the STEREO-B maximum times were 3--7 minutes later than the STEREO-A times. The time differences are probably due to both the longer travel time to STEREO-B and differences in the relative timing of the peak intensities. We were able to estimate the source locations for frequencies from 425 kHz to 1075 kHz, because the type II enhancement was observed at this frequency range at both spacecraft.  The data points in Fig.~\ref{fig2} show the source locations projected onto the ecliptic plane. The data point of the 525 kHz frequency is the only clear outlier, because the STEREO-B/WAVES (SWAVES-B) signal was too weak at this frequency to provide a reasonable estimate of the source direction. From Fig.~\ref{fig2} we can see that the radio sources were located near the Sun-STEREO-A line at W137$^{\circ}$. The mean and median longitude of the valid source locations are 122$^{\circ}$ and 126$^{\circ}$, respectively. This region corresponds the eastern side of the primary CME about 60$^{\circ}$ away from the longitude of the nose region at W70$^{\circ}$ \citep[see][]{dingetal14,chengetal14}. The location of the radio source appears reasonable and expected, because the STEREO-B spacecraft observed the CMEs above the north-east limb of the Sun as seen from the spacecraft. Therefore, the radio source region will be located towards that direction. However, it is possible that the DF source locations as seen by the STEREO-B have been shifted away from the Sun due to scattering effects. \citet{reinerstone90} have shown in their study of kilometric behind-the-limb type III radio bursts that the observed DF source locations appear to be far away from the expected location along the spiral field line. Most likely the uncertainty of the STEREO triangulation results due to the scattering effects is larger at lower frequencies.

\subsection{Wind/WAVES Observations}
In Figure~\ref{fig3} we have over-plotted on the Wind/WAVES dynamic spectrum data points showing the latitudinal location of the radio source as detected by Wind/WAVES together with the DF results of Wind/WAVES and STEREO-A/WAVES (SWAVES-A) at matching frequencies of 624 kHz and 625 kHz, respectively. The black (white) data points correspond south (north) of ecliptic direction in Fig.~\ref{fig3}(a). The normalized intensity (Int) of the radio emission observed by Wind/WAVES and STEREO-A/SWAVES are shown respectively in the panels (b) and (e) of Fig.~\ref{fig3}, the panels (c) and (f) show the estimated elevation (E) of the radio source relative to the ecliptic so that positive values  correspond to north, and the panels (d) and (g) show the azimuthal angle along the ecliptic plane relative to the Sun-spacecraft line so that positive values correspond to directions toward the west (right). Note that the angle between the ecliptic north and the solar north varies between -7.25$^{\circ}$ and +7.25$^{\circ}$ during spacecraft orbit. From panel (c) it is clear that  between 15:00 UT and 15:50 UT the source latitude at 624 kHz appears to shift from north of the ecliptic to south of the ecliptic and then return back briefly before the end of the DH type II burst. The azimuth angle also shows a minor shift towards directions closer to the Sun during the periods of northern elevation. In the panel (a) we see that similar shifts in the source latitude also occurred at other frequencies, expect at the lowest frequency of 428 kHz, where the source latitude stayed north of the ecliptic. There appears to be two main source regions that alternately dominate the observed type II emission at both spacecraft. It seems that at most frequencies the first phase of the burst was dominated by a source more towards the north and the Sun. This source direction seems to reappear after 15:50 UT, just before the end of the enhancement of the DH type II emission.

If we compare the Wind/WAVES observations in the panels (b)--(d) with those of the SWAVES-A shown in the panels (e)--(g), we see that the time evolution of the SWAVES-A profiles are quite similar to those of the WAVES. The changes of the source latitude in the SWAVES-A data are not as large as in the WAVES data. Therefore, it is likely that the dominant fraction of the DH type II emission originated from the same source region, even though the Wind and STEREO-A spacecraft were located at the opposite side of the CMEs. Even the SWAVES-B observations shows similar general time profiles, even though SWAVES-B was able to detect only the most intense part of the DH type II burst. \citet{thejappaetal07} have shown that scattering of the radio waves increases the angular visibility of the type II burst.

We also estimated the source location using the two matching pairs of frequencies (428/425 kHz and 624/625 kHz) of WAVES and SWAVES-A by calculating the closest points on the lines corresponding to the WAVES and SWAVES-A source directions. We selected two peaks that were visible at both spacecraft. The selected peaks can be seen in the panels (b) and (e) of Fig.~\ref{fig3} around 15:00--15:05 UT (first peak) and 15:12--16:22 UT (second peak), respectively. Table~\ref{tbl1} lists the longitude ($L$), the ecliptic ($R_{ecl}$) and radial ($R_{rad}$) distance from the Sun for each closest point. Also the distance ($\Delta R$) between the WAVES and SWAVES-A closest points is given in Table~\ref{tbl1}. The ecliptic distances and longitudes were in the range of 0.25--0.32 AU and 123$^{\circ}$--126$^{\circ}$, respectively, except for the first peak observed at 428 kHz and 425 kHz. In this case the ecliptic distances of the WAVES (0.10 AU) and SWAVES-A (0.15 AU) closest points are unusually close to the Sun and also the distance between the closest points is quite large (0.079 AU). The other source locations corresponded better to those obtained from the STEREO triangulation shown in Fig.~\ref{fig2}. However, the location of 428/425 kHz source is closer to the Sun that that obtained from the STEREO triangulation. This could indicate that scattering of the radio emission shifted the STEREO-B DF locations away from the Sun. The triangulation results of the WAVES and SWAVES-A DF measurements could have a higher degree of uncertainty, because the CMEs were propagating between the Wind and STEREO-A spacecraft. Overall similarity of the time profiles of the DF measurements in Fig.~\ref{fig3} indicates that the type II emission detected  at both spacecraft originated from the same radio source. However, if the radio source was spatially widely extended, it is possible that the type II emission recorded at each spacecraft was dominated by different sections of the radio source. Although the minimum time resolution (43 seconds) of the DF measurements is quite low due to frequency sweeping receivers, cross correlation analysis indicates that the WAVES detected the emission at 624 kHz $\sim$ 1 min and at 428 kHz $\sim$ 2 min later than the SWAVES-A (see panels (b) and (e) in Fig.~\ref{fig3}). Therefore, the radio source of type II emission probably was located closer to the STEREO-A. Here we have also assumed that travel time differences between fundamental and harmonic emission due to scattering are sufficiently small to be ignored \citep[see, e.g.,][]{hoangetal98}.

In Figure~\ref{fig4} we have over-plotted the estimated WIND/WAVES source directions on the two LASCO/C3 running difference images at 15:06 UT and 15:30 UT. The directions are 5-min averages centered at the maximum of the intensity. The DF method can resolve the source direction more accurately at the maximum intensity, where the signal-to-noise ratio is high \citep{cecconizarka05}. The times of the eight maxima are 14:45 UT (1040 kHz), 14:48 UT (916 kHz), 14:50 UT (804 kHz), 15:09 UT (708 kHz), 15:18 UT (624 kHz), 15:40 UT (548 kHz), 15:26 UT (484 kHz) and 15:38 UT (428 kHz). Clearly the source directions at the lower frequencies below $\sim$ 800 kHz suffer from a shift away from the Sun due to scattering of the radio waves in the IP medium \citep[see, e.g.,][]{thejappaetal07}. It is a well-known fact that the scattering increases towards the lower frequencies. However, waves at the three highest frequencies scatter less and source directions are near the leading edge of the primary CME in the 15:06 UT LASCO/C3 image indicating that the most likely radio source is the CME-CME interaction. From the position angle of the source directions at 548 kHz and 624 kHz and from Fig.~\ref{fig3}(a) we see that at these frequencies the maximum intensity originated from the source south of the ecliptic. At the 708 kHz the elevation was near the ecliptic at the time of the maximum. Therefore, the more southern radio source is most likely located at the opposite side of the nose region of the shock driven by the primary CME.

\section{Summary and Discussion}
\begin{figure}
\epsscale{.50}
\plotone{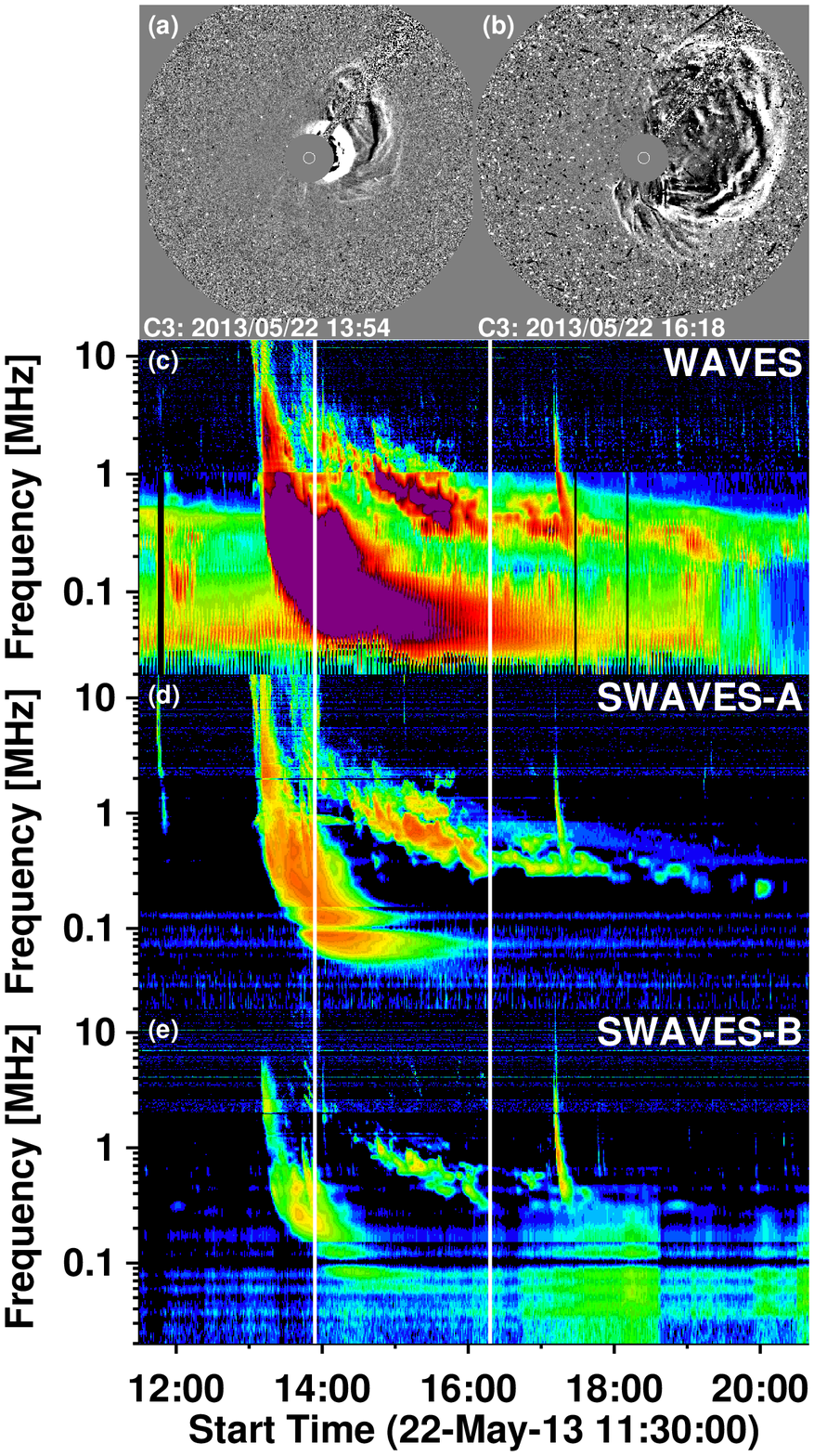}
\caption{(top) LASCO/C3 running difference images showing the two CMEs on 2013 May 22 at (a) 13:54 UT and (b) 16:18 UT. The white circle plotted on the gray occulting disk at the center of the image depicts the solar disk. The three panels below the LASCO images show (c) Wind/WAVES, (d) STEREO-A/WAVES, and (e) STEREO-B/WAVES radio dynamic spectra. The white vertical lines mark the times of the LASCO/C3 images and they approximately delimit the main enhancement of the DH type II emission due to the CME-CME interaction. The DH type II burst was preceded by a strong type III burst starting around 13:10 UT. See the electronic edition of the Journal for a color version
of this figure.\label{fig1}}
\end{figure}

We used DF analysis of Wind/WAVES and STEREO/WAVES observations to study the radio source directions during an enhancement of a DH type II radio burst on 2013 May 22, which coincided with a CME-CME interaction visible in the field of view (FOV) of LASCO/C3 coronagraph (see Fig.~\ref{fig1}). CMEs will interact when the slower CME launched earlier from the same source region is overtaken by the faster primary CME. This connection between the enhancement of the DH type II radio emission and interacting CMEs was first suggested by \citet{gopalswamyetal01,gopalswamyetal02a}. The \citet{gopalswamyetal02a} results were based on spatial information in white light and temporal information in radio. Here we confirm the relationship using spatial information at both wavelengths. On 2013 May 22 the preceding CME was first observed in the FOV of the LASCO/C2 coronagraph at 08:48 UT. Its source region was at N17$^{\circ}$W62$^{\circ}$ (C2.3 flare) and the sky-plane speed of the the CME was 687 km~s$^{-1}$. The primary CME with a sky-plane speed of 1466 km~s$^{-1}$ was detected at 13:08 UT by LASCO/C2. The estimated source location was at N15$^{\circ}$W70$^{\circ}$ (M5.0 flare). In addition, the GCS model fitting by \citet{dingetal14} shows that the propagation directions of the preceding and primary CME differed only in the heliographic latitude. The primary CME propagated closer to the solar equator (heliographic latitude $\sim$ 13$^\circ$) than the preceding CME (heliographic latitude $\sim$ 30$^{\circ}$). The start of the DH type II enhancement around 13:35 UT was observed by STEREO/WAVES and Wind/WAVES radio instruments. This radio burst intensified significantly about an hour later around 1 MHz and the intense type II emission continued  until $\sim$ 16:20 UT down to the  $\sim$ 300 MHz frequency level.

\begin{figure}
\epsscale{.80}
\plotone{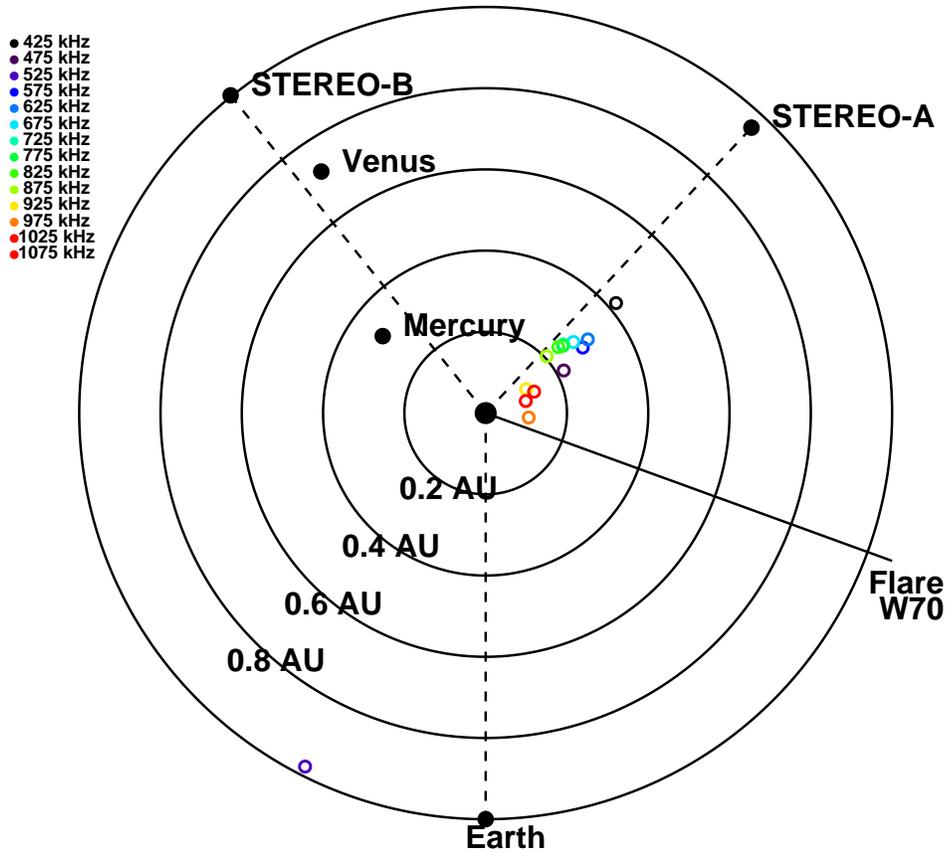}
\caption{Radio source locations obtained usind direction finding analysis and triangulation on the STEREO-A and B data. The source locations are projected onto the ecliptic plane. The concentric circles give the radial distance from the Sun. The heliographic longitude of the associated flare and the locations of the inner planets are also marked. See the electronic edition of the Journal for a color version of this figure.\label{fig2}}
\end{figure}

During the 2013 May 22 solar eruptions the STEREO-A and B spacecraft were located at the heliographic longitudes of W137$^{\circ}$ and E141$^{\circ}$, respectively. By triangulating of the radio source directions obtained using DF analysis of data from both STEREO spacecraft, we found that the projections of radio sources to the ecliptic plane at the frequency range of 425--1075 kHz were located near the Sun-STEREO-A line, about 60$^{\circ}$ east of W70$^{\circ}$, the heliographic longitude of the nose of the shock driven by the primary CME (see Fig.~\ref{fig2}). The determination of the the radio source location might have been affected by the fact that the dominant radio source might not have been directly visible at the STEREO-B spacecraft, because the spacecraft viewed the event from behind. As seen from the STEREO-B spacecraft, the CMEs were launched behind the Sun and the CME interaction was visible above the north-east limb of the Sun. However, the actual magnitude of this effect is difficult to estimate. It is possible that scattering causes the longitudinal angles of the radio source observed at STEREO-B to shift to the east, this would artificially elongate the trajectory. \citet{reinerstone90} have studied kilometric type III radio events from behind-the-limb sources. They observed that the DF source locations of the kilometric type III bursts did not follow the spiral field line as expected, instead the DF locations of the type III source at lower frequencies were far away from the expected location. They suggested that the direct radiation from the type III source could not reach the spacecraft, because obstructing regions of plasma with the plasma frequency matching the frequency of the measured radio emission were located between the source and the spacecraft. Therefore, the radio emission reaching the spacecraft must have been scattered around the obscuring plasma region and the arrival direction of the observed emission shifted far away from the spacecraft--Sun line.

\begin{figure}
\epsscale{0.55}
\plotone{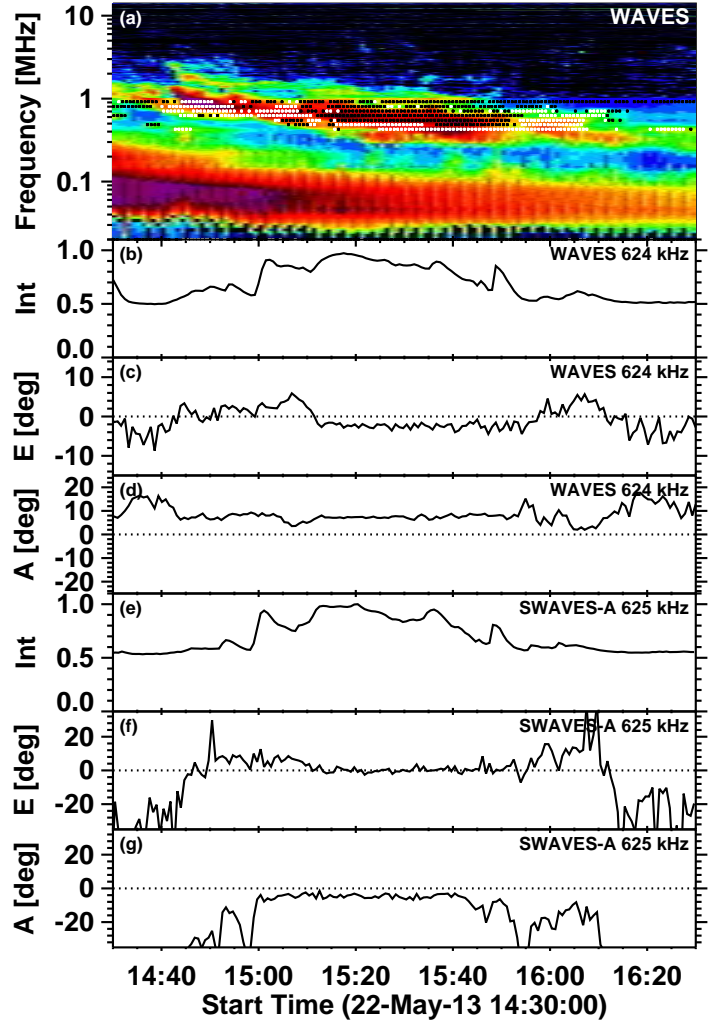}
\caption{(a) Wind/WAVES radio dynamic spectra overplotted by data points that indicate the radio source latitudinal locations for each DF frequency. The black (white) points correspond source location south (north) of ecliptic. The bright feature visible below the DH type II burst is the low-frequency section of the strong type III burst. The panels below the dynamic spectrum show the normalized intensity (Int), the elevation (E) above the ecliptic and the azimuth (A) relative to the Sun-spaceraft line obtained from the DF analysis of the Wind/WAVES (panels (b)--(d)) and STEREO-A/SWAVES (panels (e)--(g)) data at the matching frequencies of 626 kHz and 625 kHz, respectively. See the electronic edition of the Journal for a color version of this figure.\label{fig3}}
\end{figure}

\begin{figure}
\epsscale{1.00}
\plotone{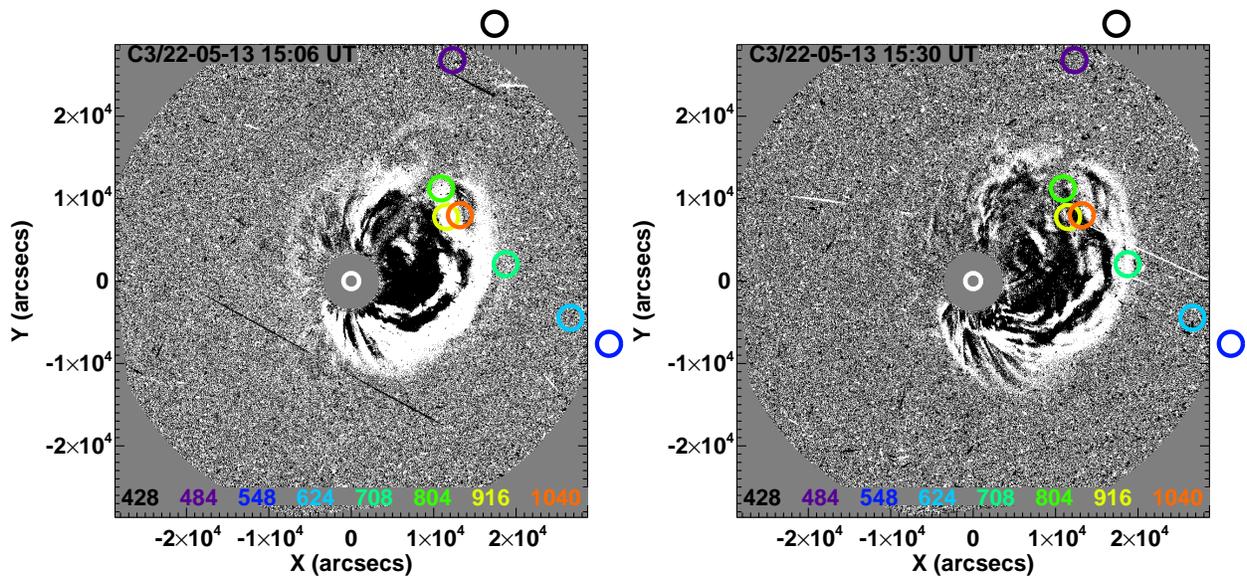}
\caption{The DF source directions (circles) estimated from the Wind/WAVES data overplotted on the LASCO/C3 running difference images at 15:06 UT and 15:30 UT. The colored numbers at the bottom indicate the corresponding frequencies in kHz. The white circle plotted on the gray occulting disk at the center of the image depicts the solar disk. See the electronic edition of the Journal for a color version of this figure.\label{fig4}}
\end{figure}

In addition, we studied the radio source direction using the Wind/WAVES data. The data revealed clear shifts in the latitude during the type II burst indicating that the source location of the dominant type II emission changed during the type II enhancement (Fig.~\ref{fig3}). The source region appeared first to the north of the ecliptic and then shifted to south of the ecliptic at most studied frequencies. The comparison of the WAVES source latitude at 624 kHz to the SWAVES-A source latitude at 625 kHz indicated similar shifts in the SWAVES-A data. When we over-plotted the locations averaged over 5-minutes at the time of the maximum intensity on the white-light images of the LASCO/C3 coronagraph, the source locations varied. It was clear that scattering of radio waves during the propagation from the source region to the observing spacecraft shifted the locations away from the Sun at frequencies below $\sim$ 800 kHz. Because scattering increases when the frequency decreases, the source direction at higher frequencies should be less affected. Also our triangulation of the Wind and STEREO-A DF measurements at two matching frequency pairs found the type II source at the lower frequency pair of 428/425 kHz was closer to the Sun than that of estimated using STEREO-A and B DF measurements, another likely indication of scattering effects. The comparison of the Wind DF locations with the LASCO white-light images showed that the type II radio source locations at 804 kHz, 916 kHz and 1040 kHz were near the leading edge of the primary CME towards the latitudinal propagation direction of the preceding CME at N30$^{\circ}$ as estimated by \citet{dingetal14}. At the frequencies of 548 kHz and 624 kHz the maximum intensity originated from the southern source region. \citet{dingetal14} estimated that the latitude of the nose region of the primary CME was at N13$^{\circ}$ \citep[see also ][]{chengetal14}, therefore the southern source was latitudinally at the opposite side of the nose of the primary CME. Finally, our analysis
confirms the results of an earlier study by \citet{martinezoliverosetal12b}, where they used DF analysis to show that radio emission during the 2010 August 1 type II burst was associated with the concurrent CME-CME interaction.

In summary, we found that radio source directions determined from the Wind/WAVES DF measurements around 1 MHz correspond the location of the leading edge of the primary CME seen in the white-light images of the LASCO/C3 coronagraph. This correspondence at both wavelengths confirms that the enhancement of type II emission originates from the CME-CME interaction region, as suggested by \citet{gopalswamyetal01}. However, the latitudinal location of the type II radio source shifted during the radio enhancement indicating that the spatial location of the dominant source of type II emission varies as CMEs interact. We also noticed that at frequencies below 800 kHz the radio locations appeared to be shifted away from the CME leading edge and the Sun confirming that at these lower frequencies scattering of radio waves significantly affects their propagation.

\acknowledgments

We are grateful to J. C. Mart{\'{\i}}nez Oliveros for providing his version of the STEREO/WAVES direction finding code. SOHO is an international cooperation project between ESA and NASA. P.M.\ and S.A.\ were partially supported by NSF grant AGS-1358274. V.K. thanks the support of the Praemium Academiae award of the Czech Academy of Sciences and the Czech Science Foundation grant GAP209/12/2394.

\clearpage

\clearpage

\begin{deluxetable}{lcccccccccccccc}
\tabletypesize{\scriptsize}
\tablecaption{Results from WAVES and SWAVES-A triangulation.\label{tbl1}}
\tablewidth{0pt}
\tablehead{
\colhead{ } & \multicolumn{2}{c}{WAVES(428 kHz)} & \colhead{ } & \multicolumn{2}{c}{SWAVES-A(425 kHz)} & \colhead{ } & \multicolumn{2}{c}{WAVES(624 kHz)} & \colhead{ } & \multicolumn{2}{c}{SWAVES-A(625 kHz)} & \\  \cline{2-3} \cline{5-6} \cline{8-9} \cline{11-12}
\colhead{Peak\#} & \colhead{$L$} & \colhead{$R_{ecl}(R_{rad}$)} & \colhead{ } & \colhead{$L$} & \colhead{$R_{ecl}(R_{rad}$)} & \colhead{$\Delta R$} & \colhead{$L$} & \colhead{$R_{ecl}(R_{rad}$)} & \colhead{ } & \colhead{$L$} & \colhead{$R_{ecl}(R_{rad}$)} & \colhead{$\Delta R$} \\
\colhead{ } & \colhead{(deg)} & \colhead{(AU)} & \colhead{ } & \colhead{(deg)} & \colhead{(AU)} & \colhead{(AU)} & \colhead{(deg)} & \colhead{(AU)} & \colhead{ } & \colhead{(deg)} & \colhead{(AU)} & \colhead{(AU)}
}
\startdata
1st Peak &83 &0.10(0.17) & &123 &0.28(0.28) & 0.079 &80 &0.15(0.25) & &124 &0.28(0.28) & 0.009 \\
2nd Peak &125 & 0.31(0.32) & &126 &0.31(0.32) &0.026 &124 &0.25(0.25) & &124 &0.25(0.25) & 0.008 \\
\enddata
\end{deluxetable}

\end{document}